\documentclass[aps,prl,showpacs,showkeys,twocolumn]{revtex4-1}
\usepackage{graphicx,hyperref,amsmath,amsfonts,comment}
\usepackage{epstopdf,color,bm,multirow,rotating,soul}
\usepackage[T2A]{fontenc} \usepackage[utf8]{inputenc}
\usepackage[russian, english]{babel}
\begin{document}
\title[Robust diagnostics of dark counts for quantum networks]{
Robust diagnostics of dark counts for quantum networks}
\author{Nikolay Sergeevich Perminov$^{1,2,3,*}$,
Maxim Aleksandrovich Smirnov$^{1}$,
Konstantin Sergeevich Melnik$^{1,2}$,\\
Lenar Rishatovich Gilyazov$^{1,2}$,
Oleg Igorevich Bannik$^{1,2}$,
Marat Rinatovich Amirhanov$^{2}$,\\
Diana Yurevna Tarankova$^{4}$,
and
Aleksandr Alekseevich Litvinov$^{1,2}$}
\affiliation{$^{1}$ Kazan Quantum Center, Kazan National Research Technical University n.a. A.N.Tupolev-KAI, 10 K. Marx, Kazan 420111, Russia}
\affiliation{$^{2}$ "KAZAN QUANTUM COMMUNICATION" LLC, 10 K. Marx, Kazan 420111, Russia}
\affiliation{$^{3}$ Zavoisky Physical-Technical Institute, Kazan Scientific Center of the Russian Academy of Sciences, 10/7 Sibirsky Tract, Kazan 420029, Russia}
\affiliation{$^{4}$ Department of Radio-Electronics and Information-Measuring Technique, Kazan National Research Technical University n.a. A.N.Tupolev-KAI, 10 K. Marx, Kazan 420111, Russia}
\email{qm.kzn@ya.ru}
\begin{abstract}
In this work, we study the timestamps in the registration of single-photon detector counts in quantum communications. The analysis of afterpulse counts on the basis of several approaches is carried out. Robust diagnostics of keys is proposed to increase the security of quantum networks.
\end{abstract}


\maketitle

\section{Noise diagnostics}
Robust diagnostics of the working complexes of quantum communication (QC) \cite{vtyurina2018} is crucial for the implementation of quantum networks capable of working in both urban \cite{bannik2017multinode,news1} and trunk \cite{bannik2019noise,news2} standard fiber-optic communication lines. At the same time, from the point of view of communication systems, trunk quantum networks (TQN) with lines of more than 100 km and losses between nodes of more than 25 dB, where the signal-to-noise ratio cannot be considered large, are particularly difficult to implement \cite{litvinov2019visual}. From the point of view of fundamental statistics, the fundamental difficulty here lies in the fact that despite the relatively low average percentage of errors, the magnitude of the error span and the variance of errors can be extremely large, which entails low reliability of diagnosing errors when performing continuous tests of the TQN and especially large-scale TQN with a large number of nodes.

The necessary solution for a similar problem in communication theory is the use of prognostic monitoring tools and error filtering, which will increase the reliability of security parameters for inter-city QC in continuous use. We also note that due to large errors for the limit passport operating modes of QC complexes, the security of the communication complex should be determined by a special diagnostic system that is different from the diagnostic system for operating passport modes. This difference in diagnosis is also due to the fact that the reliability of error determination in the limiting mode effectively depends on a significantly larger number of physical factors, which are not always easy to track within the framework of a single structurally finished product. That is, a more powerful specialized diagnostic system should be delivered with the QC complex as a separate product and be able to work in the background of the TQN to track the entire history of changes in the line performance even with the QC complex in the node disabled. In addition, such a system should be able to track additional performance factors associated with the statistics of quantum and classical noise.

In large-scale TQN, according to the theory of reliability, control requirements should be higher than for conventional QC due to the large number of elements and a high level of noise that cannot be eliminated for long-range QC. A significant part of the noise in the TQN is due to various factors arising from the registration of single photons by highly sensitive single-photon detectors. Therefore, explicit statistical accounting of the noise of quantum detectors allows you to most correctly select the mode of use of the detectors to realize the most efficient quantum coupling with the highest signal to noise ratio. Direct statistical analysis and robust diagnostics of the noise of quantum detectors can be performed by distributing the time stamps of quantum keys, which are available for the online diagnostic system and, in our opinion, carry a significant amount of information about the QC operability and the level of quantum-classical security of QC complexes as a whole.

In this work, we study timestamps when registering counts of single-photon detectors in QC. Post-pulse counts are analyzed based on several approaches. The conclusion is made about the proportion of dark noise and post-pulse counts in the total noise, and the limits of applicability of the theory are shown using a sequence of the ranged amplitudes. We offer non-parametric robust diagnostic of times tags in keys to increase the security of quantum networks, and also discuss the prospects of commercializing quantum-classical cloud-based security services.

\section*{Modern diagnostic methods}
One of the modern statistical diagnostic methods is a sequence of the ranged amplitudes (SRA), which is a sequence of numbers obtained from the original sequence by ordering numbers in descending (or ascending) order of their values. SRA are often used in histogram construction algorithms. The SRA sequences themselves are practically not analyzed, although they are very interesting, since when constructing them, there is no loss of information about specific values of the source data, as, for example, using the histogram method. It also gives the advantage of using much less data for statistical analysis. Analytical analysis of SRA is the essence of the so-called SRA method \cite{nigmatullin2003fluctuation,nigmatullin2010new,baleanu2010new}. When using the SRA method, the dependence of the values of the elements in the SRA sequence on their serial number (index) in the sequence is analyzed.

To search for an analytical expression, it is convenient to use an approximate expression connecting the SRA with the empirical distribution function $F(x)$ \cite{smirnov2018sequences,perminov2018comparison}:
\begin{align}\label{EQ1}
F(x,x_n)=(N+1-n(x_n))/N,    
\end{align}
where is the total number of points in the sample. In some cases, knowing the analytical form for the distribution function $F(x)$, one can find an analytical expression for the SRA of the form that can be actively used in the SRA method to quickly find the statistical parameters of the initial sample. Next, we consider two versions of expressions for SRA that describe random readings of a single-photon detector. The first one considers samples caused only by random Poisson sources (the probability of post-pulse counts is zero). The second one considers counts caused by post-pulse counts.

\section*{Dark counts of detectors}
Light sources, as well as dark noise, corresponding to Poisson processes, correspond to the probability density for the time intervals between samples, which is approximately described by a decaying exponent \cite{wiechers2016systematic,horoshko2017afterpulsing}:
\begin{align}\label{EQ2}
dF/dx=\rho(x_n)= \lambda exp(-\lambda x_n),
\end{align}
where $\lambda$ is the average frequency of avalanche events, and $x_n$ are the time intervals between samples. From (\ref{EQ1}) and (\ref{EQ2}) can be obtained in form \cite{smirnov2018sequences,horoshko2017afterpulsing}:
\begin{align}\label{EQ3}
x_n= \lambda^{-1} Ln(N/(n-1).
\end{align}
In gallium arsenide avalanche photodiodes used in QC quantum detectors, the exponential model for dark counts is violated due to post-pulse counts. The effect of post-pulse counts is the process of re-emission of charges captured by the avalanche diode traps during the previous avalanche event. The traps inside the diode are not due to the ideality of the detector and its manufacturing technology.

The first models describing post-pulse counts were constructed on the basis of a simple exponential dependence of the probability density on the inter-pulse interval \cite{itzler2012power,humer2015simple}. This expression depends on the amplitude of the probability and the decay time. However, this expression describes well the processes in the "free-run" mode and does not accurately describe the readings of devices operating in the "gating" mode. In \cite{itzler2012power}, it was shown that in this mode, post-pulse counts are well described by a power function of the form
\begin{align}\label{EQ4}
P= C t^{\alpha},
\end{align}
where $C$ and $\alpha$ are positive parameters, $t$ is time in counts. In this paper, it is assumed that a possible reason for this dependence is the wide distribution of trap energy in the used semiconductor avalanche diode. Moreover, the values of the parameters $C$ and $\alpha$ depend on the shape of the density function of the energy distribution of the traps. The characteristic value of the parameter $\alpha$ for InGaAs/InP detectors
is 1.2 $\pm$ 0.2 \cite{itzler2012power}.

Expression (\ref{EQ4}) describes with high accuracy the statistics of post-pulse readings. The reasons for this are that the charge traps in the avalanche photodiode have a wide energy distribution. This circumstance does not allow us to describe the statistics of readings by a single decaying exponent, which determines a single Arrhenia relaxation process. Of course, it is possible to improve the degree of fitting by choosing as the fitting function the sum of exponentials with different decay times, however, the number of exponents is not a fixed parameter and the physical meaning of the parameters obtained is not clear. It is important to note that each value of the probability distribution function in (\ref{EQ2}) and (\ref{EQ4}) is obtained by calculating a statistically suitable number of experimental points in a small time interval. Thus, to obtain the dependence of the probability distribution function that is suitable for analysis on the inter-pulse intervals, a large number of experimental values are required (of the order of $10^6$ values \cite{humer2015simple}).

\begin{figure}[t]
\includegraphics[width = 0.48\textwidth]{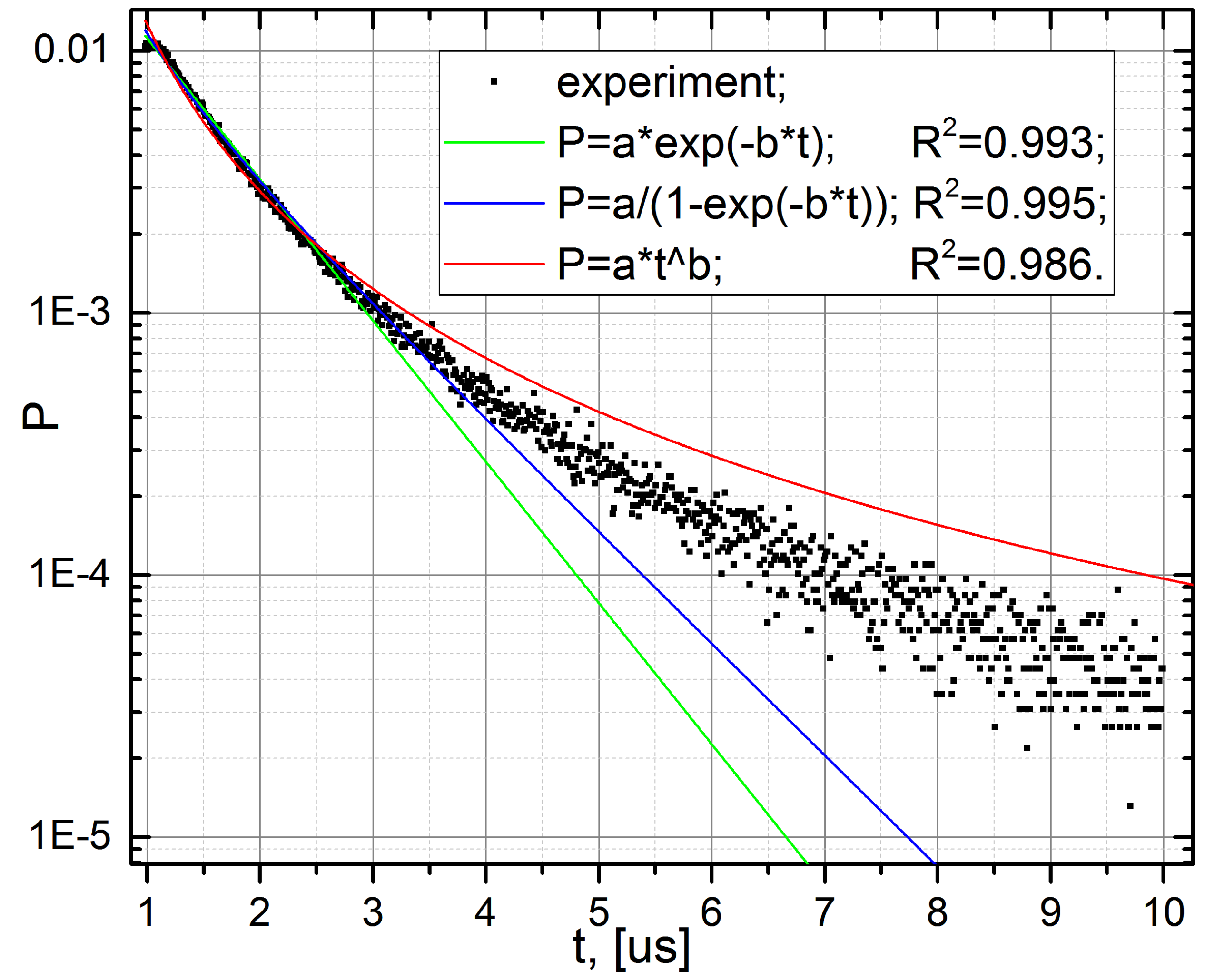}
\caption{Noise parametrization according to Itzler's theory (\ref{EQ4}) (red line), Poisson (green line) and the new theory (blue line). Here, the comparison is at the level of the empirical probability density of dark counts for the working QC system.}
\label{FIG1}
\vspace{-0.5cm}
\end{figure}

For the data obtained in the experiment, we test statistical hypotheses of the form (\ref{EQ2}), (\ref{EQ4}) with an additional factor introduced for testing flexibility. In addition, we are testing a new hypothesis for a probability of the form $P = A/(1-exp(-B t))$, which is a solution of a nonlinear equation of the form $dP/dt = a P + b P^2$, potentially corresponding to possible strong nonlinear effects on microscopic level of description of quantum detectors. The results of quantitative parametrization of dark samples with a small number of post-pulse counts for a working QC system are shown in Fig. \ref{FIG1}.

We see that for the new hypothesis and Poisson theory \cite{perminov2018comparison}, the results of estimating noise in the region of small times t<3 us are more accurate than for Itzler's theory \cite{itzler2012power}. However, the asymptotic behavior of the curves on a logarithmic scale indicates the need for a more competent account of the distribution tails in the region of large times t>3 us, which is not so simple to do in the framework of one consistent physical theory \cite{itzler2012power}. Here we see the imperfection of theoretical models, which is difficult to overcome within the framework of theoretical models of noise analysis with a small finite number of parameters, but it is easy to overcome within the framework of nonparametric statistical criteria for noise analysis, such as SRA.

\section*{SRA analysis of dark noise}
\begin{figure}[t]
\includegraphics[width = 0.48\textwidth]{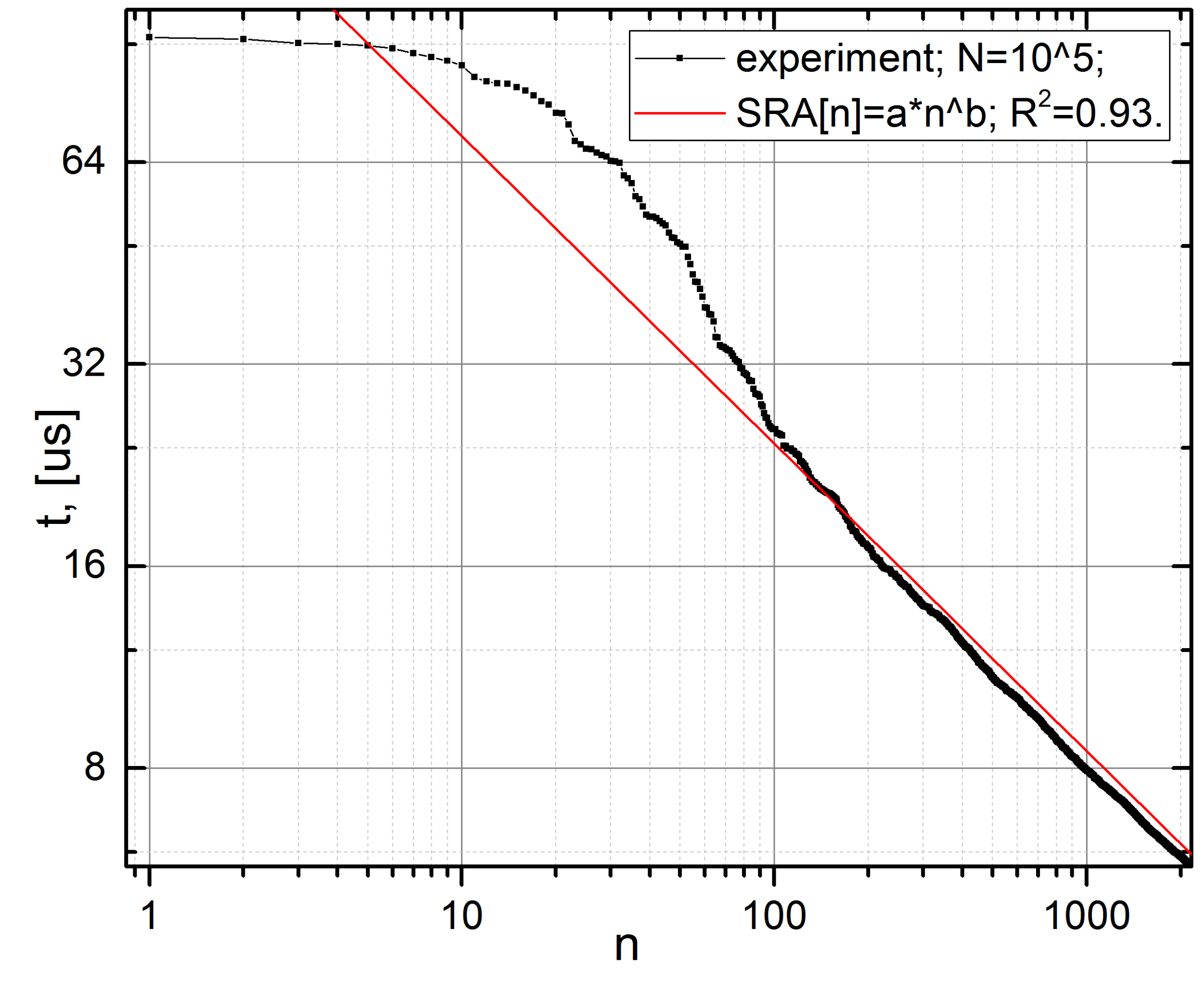}
\caption{Robust parametrization of dark photocounts using SRA $x_n$. For t>24us, Itzler's theory does not work well. Axes are given on a logarithmic scale.}
\label{FIG2}
\vspace{-0.5cm}
\end{figure}

Expression (\ref{EQ4}) we will base the model of post-pulse counts based on SRA. In the case when the samples dominated by the counts due to the effect of post-pulses, we can substitute (\ref{EQ4}) in (\ref{EQ1}) and get:
\begin{align}\label{EQ5}
x_n= x_{min} (N/(n-1))^{1/(\alpha-1)},
\end{align}
where $x_{min}$ is the minimum time value equal to the dead time $T_{ho}$, N is the number of points in the sample. Eq. (\ref{EQ5}) has only one adjustable parameter $\alpha$.
Note that Eq. (\ref{EQ3}) and (\ref{EQ5}) contain each experimental value. Thus, in this case, to obtain a dependence suitable for analysis, a much smaller number of experimental values (of the order of $10^3$–$10^4$) are required in comparison with the probability distribution functions obtained from the histograms. In fig. 2 shows the parametrization of dark noise with a sufficient proportion of post-pulse samples based on formula (\ref{EQ5}) with an additional factor for testing flexibility, which corresponds to Itzler's theory \cite{itzler2012power}. In fig. \ref{FIG2} on a logarithmic scale, we clearly see that starting from t> 24 us, Itzler's theory ceases to work. Thus, in addition to fast parametrization of noise by theoretical models, we can also quantify the applicability limits of models using non-parametric analysis of the tails of the distribution of SRA.

\section*{Relative SRA for different operating modes of QC}
Explicit statistical accounting of the noise of quantum detectors allows you to most correctly select the mode of use of the detectors to realize the most efficient quantum communication with the highest signal to noise ratio. Direct statistical analysis and robust diagnostics of the noise of quantum detectors can be done by distributing the times tags of quantum keys that are available for the online diagnostic system and a significant amount of information about the QC performance (the amount of dark noise and post-pulse counts, line interference, etc.). Moreover, a joint statistical analysis of the dynamics of bit errors in keys and key times tags can become the new standard of quantum-classical security of quantum communication complexes as a whole due to the high predictive power of nonparametric statistical criteria \cite{nigmatullin2003fluctuation,nigmatullin2010new,baleanu2010new,smirnov2018sequences,perminov2018comparison}.

\begin{figure}[t]
\includegraphics[width = 0.48\textwidth]{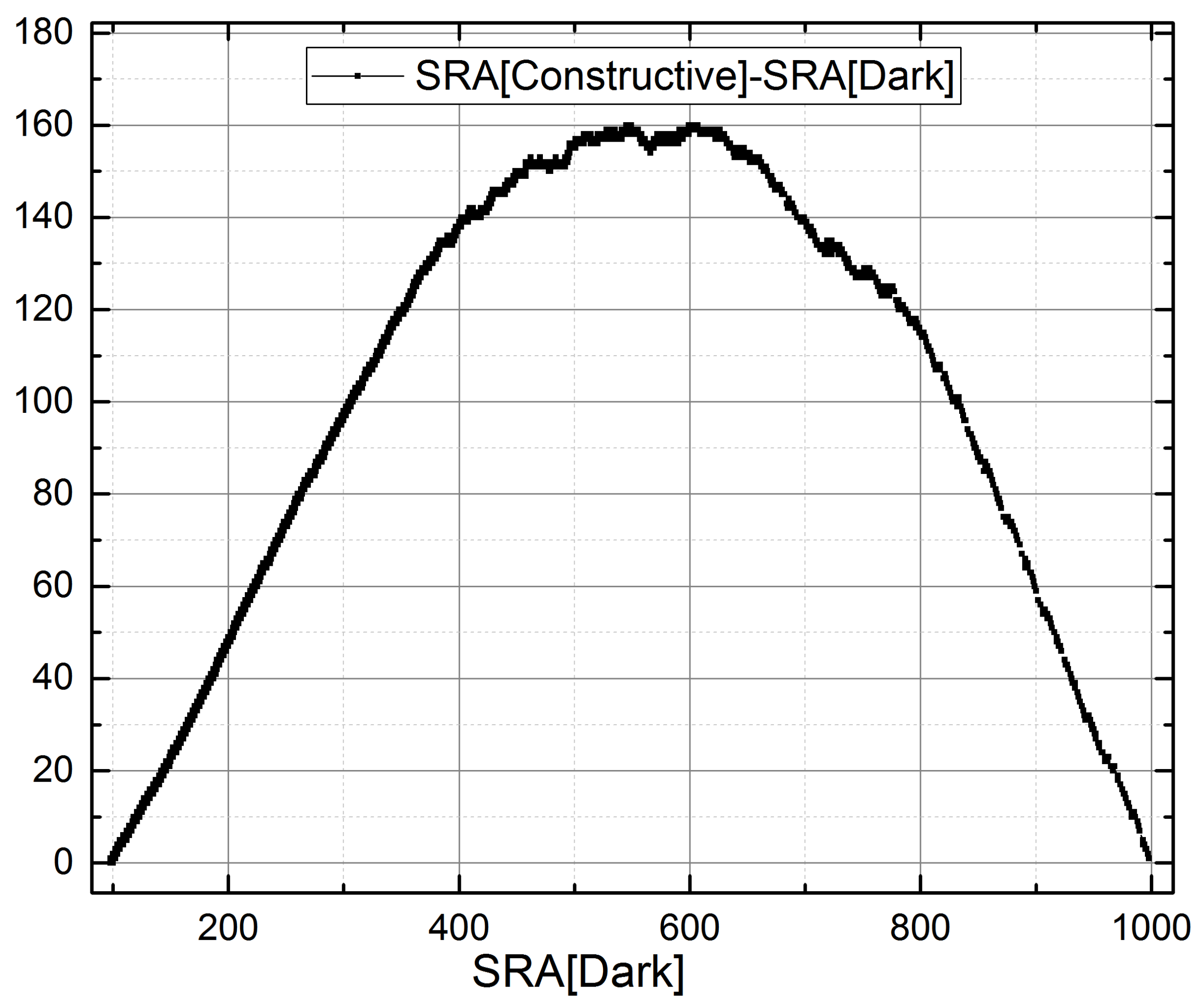}
\caption{Dependence of SRA[Constructive]-SRA[Dark] on SRA[Dark] for precise distinguishing constructive and destructive interference using times tags in keys for continuous diagnostics of changes in statistics.}
\label{FIG3}
\vspace{-0.5cm}
\end{figure}

In fig. \ref{FIG3} shows the dependence of SRA[Constructive]-SRA[Dark] on SRA[Dark] for precise distinguishing between constructive and destructive interference using times tags within in key for continuous diagnosis of changes in statistics. Unlike parametric criteria with 2-3 evaluation parameters, the criterion described here is the entire resulting dependence, that is, 1000 fitting parameters. In this sense, the power of many advanced nonparametric criteria is almost incomparable with conventional parametrization methods. Therefore, despite the difficulty in comparing nonparametric criteria with real physical factors, they are hypersensitive even with small changes in the systems under study and are able to work in the absence of a priori information. Accordingly, such robust nonparametric methods as SRA can be used as the basis for quantum-classical diagnostics of QC.

\section*{Conclusion}
Precise error diagnostics in QC depends on a large number of physical factors that are difficult to track within the framework of only one structurally finished product. In our opinion, needed an expanded robust diagnostic system, which should be supplied with the QC complex and be able to conduct a joint statistical analysis of such seemingly different parameters as the time stamps of keys and the distribution of errors within the key. Such a quantum-classical diagnostic subnet in the TQN, capable of diagnosing even small noises in the network in the absence of a priori information about the type of interference, also opens up prospects for the commercialization of quantum-classical cloud services for robust information protection.

\section*{Acknowledgments}
The team of authors expresses special gratitude to co-author M. A. Smirnov for the initial idea of a microscopic approach to the dynamics of post-pulse counts and the basic statistical verification of this hypothesis. Research in the field of statistics of quantum detectors and quantum sensors was carried out with the financial support of RFBR grant No. 19-32-80029 (basic idea – MAS) and also grant of the Government of the Russian Federation 14.Z50.31.0040, February 17, 2017 (experiment and nonparametric analysis –- NSP, MAS, KSM, LRG, OIB, AAL). The work is also partially supported in the framework of the budget theme of the laboratory of Quantum Optics and Informatics of Zavoisky Physical-Technical Institute (numerical modeling in quantum informatics –- NSP, DYT).

\section{Information about authors}
\noindent
\textbf{Nikolay Sergeevich Perminov},\\
(b. 1985), in 2008 graduated the department of Theory of Relativity and Gravity of the Physics Department of Kazan Federal University in the direction of “Physics”, researcher at the Kazan Quantum Center of the KNRTU-KAI.\\
\textit{Area of interest:} optimal control, quantum informatics, statistics, software, economics.\\
\textit{E-mail:} qm.kzn@ya.ru\\
\\
\noindent
\textbf{Maxim Aleksandrovich Smirnov},\\
(b. 1990), in 2013 graduated with honors from the graduate of the Physics Institute of the Kazan Federal University in the direction "Physics", researcher at the Kazan Quantum Center of the KNRTU-KAI.\\
\textit{Area of interest:} chemistry, nonlinear optics, coherent optics, quantum communications, optoelectronics, photonics.\\
\textit{E-mail:} maxim@kazanqc.org\\
\\
\noindent
\textbf{Konstantin Sergeevich Melnik},\\
(b. 1993), in 2018 graduated from Institute of Radio Electronics and Telecommunications of Kazan National Research Technical University in the direction of "Radio Engineering", engineer at the Kazan Quantum Center of the KNRTU-KAI.\\
\textit{Area of interest:} quantum communications, optoelectronics, photonics.\\
\textit{E-mail:} mkostyk93@mail.ru\\
\\
\noindent
\textbf{Lenar Rishatovich Gilyazov},\\
(b. 1985), in 2008 graduated from the the Physics Department of Kazan Federal University, researcher at the Kazan Quantum Center of the KNRTU-KAI.\\
\textit{Area of interest:} quantum communications, optoelectronics, photonics.\\
\textit{E-mail:} lgilyazo@mail.ru\\
\\
\noindent
\textbf{Oleg Igorevich Bannik},\\
(b. 1988), in 2012 graduated from the faculty of electronics of Saint Petersburg Electrotechnical University in the direction of "Electronics and Microelectronics", researcher at the Kazan Quantum Center of the KNRTU-KAI.\\
\textit{Area of interest:} quantum communications, optoelectronics, photonics.\\
\textit{E-mail:} olegbannik@gmail.com\\
\\
\noindent
\textbf{Marat Rinatovich Amirhanov},\\
(b. 1985), in 2008 he graduated from the faculty of power engineering Kazan National Research Technological University.\\
\textit{Area of interest:} quantum communications, network technologies, automation and programming.\\
\textit{E-mail:} m.amirhanov85@gmail.com\\
\\
\noindent
\textbf{Diana Yurevna Tarankova},\\
(b. 1995), since 2017 a student at the Institute of Radio Electronics and Telecommunications of Kazan National Research Technical University.\\
\textit{Area of interest:} programming, photonics, network technologies.\\
\textit{E-mail:} tarankovadyu@ya.ru\\
\\
\noindent
\textbf{Aleksandr Alekseevich Litvinov},\\
(b. 1985), in 2008 graduated the department of Theory of Relativity and Gravity of the Physics Department of Kazan Federal University in the direction of “Physics”, engineer at the Kazan Quantum Center of the KNRTU-KAI.\\
\textit{Area of interest:} quantum communications, optoelectronics, robust methods, quantum memory, software, economics.\\
\textit{E-mail:} litvinov85@gmail.com\\

\vspace{-1.0cm}

\bibliographystyle{ieeetr}
\bibliography{RD_DC_QN}

\begin{thebibliography}{10}

\bibitem{vtyurina2018}
A.~G. Vtyurina, V.~L. Eliseev, A.~E. Zhilyaev, A.~S. Nikolaeva, V.~N. Sergeev,
  and A.~V. Urivskiy, ``On the principal decisions of the practical
  implementation of the cryptographic devices with quantum key distribution,''
  {\em Doklady TUSUR}, vol.~21, no.~2, 2018.

\bibitem{bannik2017multinode}
O.~Bannik, V.~Chistyakov, L.~Gilyazov, K.~Melnik, A.~Vasiliev, N.~Arslanov,
  A.~Gaidash, A.~Kozubov, V.~Egorov, S.~Kozlov, A.~Gleim, and S.~Moiseev,
  ``Multinode subcarrier wave quantum communication network,'' in {\em QCrypt
  2017 held at the University of Cambridge on 18-22, September 2017 in
  Cambridge, United Kingdom}, p.~Th413, Centre for Photonic Systems, University
  of Cambridge, 2017.

\bibitem{news1}
{KQC KNRTU-KAI}, {ITMO}, and {PJSC $"$Tattelecom$"$}, ``{The first multi-node
  quantum network in the CIS was successfully launched in Kazan},'' {\em
  News-center of KNRTU-KAI}, 12 may 2017.
\newblock {\textit{In Russian news:} Первая в СНГ
  многоузловая квантовая сеть успешно
  заработала в Казани. https://kai.ru/news/new?id=5931227}.

\bibitem{bannik2019noise}
O.~I. Bannik, L.~R. Gilyazov, A.~V. Gleim, N.~S. Perminov, K.~S. Melnik, N.~M.
  Arslanov, A.~A. Litvinov, A.~R. Yafarov, and S.~A. Moiseev, ``Noise-immunity
  kazan quantum line at 143 km regular fiber link,'' {\em arXiv preprint
  arXiv:1910.10011}, 2019.

\bibitem{news2}
{KQC KNRTU-KAI}, {PJSC $"$Rostelecom$"$}, and {PJSC $"$Tattelecom$"$}, ``{In
  Russia, quantum encryption has been successfully tested on the fiber optic
  with a record distance},'' {\em Press-center of PJSC $"$Rostelecom$"$}, 25
  september 2019.
\newblock {\textit{In Russian news:} В России успешно
  протестировали квантовое шифрование на
  ВОЛС с рекордным расстоянием.
  https://www.company.rt.ru/press/news/\\
  d452080/?backurl$=$/press/$\#\_$ftnref1}.

\bibitem{litvinov2019visual}
A.~A. Litvinov, E.~M. Katsevman, O.~I. Bannik, L.~R. Gilyazov, K.~S. Melnik,
  M.~R. Amirhanov, D.~Y. Tarankova, and N.~S. Perminov, ``Visual media for
  monitoring trunk quantum networks,'' {\em arXiv preprint arXiv:1912.01426},
  2019.

\bibitem{nigmatullin2003fluctuation}
R.~Nigmatullin and G.~Smith, ``Fluctuation-noise spectroscopy and a
  “universal” fitting function of amplitudes of random sequences,'' {\em
  Physica A: Statistical Mechanics and its Applications}, vol.~320,
  pp.~291--317, 2003.

\bibitem{nigmatullin2010new}
R.~R. Nigmatullin, ``New noninvasive methods for ‘reading’of random
  sequences and their applications in nanotechnology,'' in {\em New trends in
  nanotechnology and fractional calculus applications}, pp.~43--56, Springer,
  2010.

\bibitem{baleanu2010new}
D.~Baleanu, Z.~B. G{\"u}ven{\c{c}}, J.~T. Machado, {\em et~al.}, {\em New
  trends in nanotechnology and fractional calculus applications}.
\newblock Springer, 2010.

\bibitem{smirnov2018sequences}
M.~A. Smirnov, N.~S. Perminov, R.~R. Nigmatullin, A.~A. Talipov, and S.~A.
  Moiseev, ``Sequences of the ranged amplitudes as a universal method for fast
  noninvasive characterization of {SPAD} dark counts,'' {\em Applied optics},
  vol.~57, no.~1, pp.~57--61, 2018.

\bibitem{perminov2018comparison}
N.~S. Perminov, M.~A. Smirnov, R.~Nigmatullin, and S.~A. Moiseev, ``Comparison
  of the capabilities of histograms and a method of ranged amplitudes in noise
  analysis of single-photon detectors,'' {\em Computer Optics}, vol.~42, no.~2,
  pp.~338--342, 2018.

\bibitem{wiechers2016systematic}
C.~Wiechers, R.~Ram{\'\i}rez-Alarc{\'o}n, O.~R. Mu{\~n}iz-S{\'a}nchez, P.~D.
  Y{\'e}piz, A.~Arredondo-Santos, J.~G. Hirsch, and A.~B. U’Ren, ``Systematic
  afterpulsing-estimation algorithms for gated avalanche photodiodes,'' {\em
  Applied optics}, vol.~55, no.~26, pp.~7252--7264, 2016.

\bibitem{horoshko2017afterpulsing}
D.~Horoshko, V.~Chizhevsky, and S.~Y. Kilin, ``Afterpulsing model based on the
  quasi-continuous distribution of deep levels in single-photon avalanche
  diodes,'' {\em Journal of Modern Optics}, vol.~64, no.~2, pp.~191--195, 2017.

\bibitem{itzler2012power}
M.~A. Itzler, X.~Jiang, and M.~Entwistle, ``Power law temporal dependence of
  ingaas/inp spad afterpulsing,'' {\em Journal of Modern Optics}, vol.~59,
  no.~17, pp.~1472--1480, 2012.

\bibitem{humer2015simple}
G.~Humer, M.~Peev, C.~Schaeff, S.~Ramelow, M.~Stip{\v{c}}evi{\'c}, and
  R.~Ursin, ``A simple and robust method for estimating afterpulsing in single
  photon detectors,'' {\em Journal of Lightwave Technology}, vol.~33, no.~14,
  pp.~3098--3107, 2015.

\end{thebibliography}

\end{document}